%% file: 26-1-13.tex
\documentclass[conference]{IEEEtran}
\IEEEoverridecommandlockouts

\usepackage{cite}
\usepackage{amsmath,amssymb,amsfonts}
\usepackage{algorithmic}
\usepackage{graphicx}
\usepackage{textcomp}
\usepackage{xcolor}
\usepackage[table]{xcolor}

\usepackage{multirow}

\usepackage{algorithm}
\usepackage{algorithmic}
\usepackage{amsthm}

\usepackage{subcaption}
\usepackage{graphicx}
\IEEEaftertitletext{\vspace{-2\baselineskip}}
\theoremstyle{definition}
\newtheorem{example}{Example}

\usepackage{tikz}
\usepackage{pgfplots}
\usetikzlibrary{plotmarks}
\usetikzlibrary{matrix,positioning,shapes,arrows,decorations,calc,fit}
\usetikzlibrary{decorations.pathreplacing}
\usetikzlibrary{spy,backgrounds}
\usetikzlibrary{external}
\usetikzlibrary{patterns}
\usetikzlibrary{shapes,arrows}
\usetikzlibrary{shadows.blur}
\usetikzlibrary{shapes.symbols}
\usetikzlibrary{
	pgfplots.groupplots,
	matrix
}

\usepackage[
  paper=letterpaper,
  top=0.7in,
  bottom=1.02in,
  inner=0.625in,
  outer=0.65in,
  columnsep=0.21in
]{geometry}

\makeatother
\def\BibTeX{{\rm B\kern-.05em{\sc i\kern-.025em b}\kern-.08em
    T\kern-.1667em\lower.7ex\hbox{E}\kern-.125emX}}
\begin{document}

\title{LLM-Viterbi: Semantic-Aware Decoding for Convolutional Codes\\
}

\author{
\IEEEauthorblockN{Zhengtong Li, Chentao Yue, Jiafu Hao, Branka Vucetic,  and Yonghui Li}

\IEEEauthorblockA{School of Electrical and Computer Engineering, the University of Sydney, Sydney, NSW 2006\\
                 Email:  \{zhengtong.li, chentao.yue, Jiafu.hao, branka.vucetic, yonghui.li\}@sydney.edu.au}

\thanks{The work of Chentao Yue was supported by ARC DECRA under Grant DE250101332. Code available: https://github.com/Todd-6/LLM-Viterbi}
}

\maketitle
\thispagestyle{empty}
\begin{abstract}
Traditional wireless communications rely solely on bit-level channel coding for error correction, without exploiting the inherent linguistic structure of the data source. This paper proposes a large language model (LLM) Viterbi decoder that integrates LLM priors into the Viterbi decoding for text transmission over AWGN channels. The proposed decoder maintains multiple candidate paths during the Viterbi decoding and periodically evaluates path reliabilities using a fine-tuned Byte-level T5 (ByT5) language model. By combining channel reliability metrics with semantic probability from the LLM, it outputs the path that maximizes the joint likelihood of channel observations and linguistic coherence. Simulations show that our decoder achieves significant performance gains over conventional Viterbi decoding in terms of both block error rate (BLER) and semantic similarity. For convolutional codes with constraint length 3, it achieves approximately 1.5 dB more coding gain in BLER, with over 50\% improvements in semantic similarity. 
The framework can extend to other structured data sources beyond text.
\end{abstract}

\vspace{-0.8em}
\section{Introduction}
\vspace{-0.5em}

Reliable data transmission is fundamental to modern communication systems. Classically, this relies on the source-channel separation principle \cite{shannon1948mathematical}, which assumes that source coding and channel coding can be optimized independently without loss of optimality. Traditional channel coding, such as convolutional codes \cite{elias1955coding} and LDPC codes \cite{gallager1962ldpc}, perform decoding under the assumption that input information bits follow a uniform distribution. While this assumption simplifies decoder design and enables mathematical analysis, it overlooks the inherent residual redundancy in natural data sources, such as text, image, and video. 
To exploit this redundancy, classical joint source-channel decoding (JSCD) approaches \cite{hagenauer1995source} typically model source statistics as simple Markov chains and incorporate them into the decoding process. However, these low-order statistical models are insufficient for complex, long-range semantic dependencies in structured data sources.

Addressing this limitation, recent research has pivoted towards semantic communication, leveraging deep neural networks (DNNs) to capture high-dimensional source features.
A prominent paradigm in this domain is deep joint source-channel coding (Deep JSCC) \cite{xie2021deep,gunduz2022beyond,bourtsoulatze2019deep}, which employs autoencoder-based architectures to map source data directly to continuous channel symbols, learning to extract and protect semantic information end-to-end.
While demonstrating impressive empirical performance against channel noise, such "black-box" models lack rigorous algebraic structure and the mathematical optimality guarantees of classical coding theory. Moreover, these approaches abandon the well-established source-channel separation architecture \cite{shannon1948mathematical}, requiring complete redesign of the communication stack.

Recent work has integrated generative AI and large language models (LLM) into physical layer signal processing tasks\cite{huynh2024generative}, such as channel estimation, constellation design, and signal detection. In channel coding, the ``short wins long'' approach \cite{hao2025short} leverages LLMs for additional error correction. Specifically, a message is transmitted using multiple short linear block codewords, and when some blocks contain errors, correctly decoded blocks provide context for an LLM to recover erroneous blocks. 
However, due to the nature of block codes, this two-stage method suffers from fundamental limitations. Block codes must be decoded entirely before providing input to the LLM, forcing the LLM to operate independently without access to soft channel information. Additionally, LLM outputs cannot feed back to the channel decoder, as decoding has already completed. Ideally, the LLM should provide source priors $P(\mathbf{u})$ that are fully integrated into the channel decoding via
\begin{equation}
\hat{\mathbf{u}} = \arg\max_{\mathbf{u}} P(\mathbf{u}|\mathbf{y}) = \arg\max_{\mathbf{u}} P(\mathbf{y}|\mathbf{u}) \cdot P(\mathbf{u}),
\end{equation}
where $P(\mathbf{y}|\mathbf{u})$ represents the channel likelihood.

In this paper, we propose LLM-Viterbi decoding that integrates language model priors directly into the Viterbi decoding process for convolutional codes. We focus on text transmission, where linguistic structure provides powerful semantic priors. The trellis structure of convolutional codes, combined with $K$-best Viterbi decoding \cite{seshadri1994list}, maintains multiple candidate paths simultaneously.  We then use a fine-tuned ByT5 language model \cite{byt5} to periodically evaluate and prune these paths based on joint scoring that combines channel reliability with linguistic plausibility. This creates a collaborative decoder where semantic information guides decoding decisions in real-time.
Simulations over AWGN channels demonstrate significant performance gains in block error rate (BLER) and semantic similarity. For example, with rate-1/2 convolutional codes with constraint length $\nu = 3$, LLM-Viterbi achieves approximately 1.5 dB coding gain in BLER compared to standard Viterbi \cite{viterbi}. Furthermore, at 1 dB SNR, our decoder achieves SBERT \cite{sbert} score of 0.82, compared to only 0.52 for standard Viterbi. We show that the integration of LLM increases decoding latency, but this tradeoff yields improvements in both BLER and semantic preservation.


\vspace{-0.5em}
\section{Preliminaries}
\label{sec:system_model}
\vspace{-0.3em}

\subsection{System Model}
\vspace{-0.3em}
At the transmitter, a text message is first converted into a token sequence $\mathbf{t} = (t_1, t_2, \ldots, t_{L_T})$ of length $L_T$ using character-level tokenization where each $t_i$ corresponds to a single character, enabling compatibility with the ByT5 language model (see Section \ref{sec::Byt}). Each character is represented using 8-bit ASCII encoding to form the binary sequence $\mathbf{u} = (u_1, u_2, \ldots, u_L)$ of length $L = 8L_T$. Note that any other symbol-wise source coding schemes are applicable.

Then, $\mathbf{u}$ is encoded by a rate-$R$ convolutional encoder with constraint length $\nu$ and generator polynomials $g(x)$, which produces a codeword $\mathbf{c} = (c_1, c_2, \ldots, c_M)$ of length $M = L/R$. Binary phase shift keying (BPSK) modulation maps the encoded bits ($0 \to +1$ and $1 \to -1$) to the transmitted signal $\mathbf{x} = (x_1, x_2, \ldots, x_M)$.

The transmitted signal passes through an AWGN channel. The received signal is given by
\begin{equation}
\mathbf{y} = \mathbf{x} + \mathbf{n},
\end{equation}
where $\mathbf{n} = (n_1, n_2, \ldots, n_M)$ contains independent and identically distributed (i.i.d.) Gaussian noise with zero mean and variance $\sigma^2 = N_0/2$. The signal-to-noise ratio (SNR) is defined as $\text{SNR} = E_b/N_0$, where $E_b$ is the energy per information bit.

\begin{figure} [t]
     \centering
     \begin{subfigure}[b]{0.48\columnwidth}
        \centering
            \input{figures/7o5_encoder}
            \vspace{-1em}
         \caption{Encoder}

         \label{Fig::7o5::encoder}
     \end{subfigure}
     \hfill
     \begin{subfigure}[b]{0.45\columnwidth}
         \centering
            \input{figures/7o5_trellis}
             \vspace{-1em}
        \caption{Trellis}  
        \label{Fig::7o5::trellis}
     \end{subfigure}
     \vspace{-0.5em}
     \caption{The trellis and encoder of $(1, 7_{\mathrm{oct}}/5_{\mathrm{oct}})$}
      \vspace{-1.5em}
     \label{Fig::7o5::trellis&encoder}
\end{figure}

Convolutional code is defined by generator polynomials $g(x)$ \cite{viterbi}. For constraint length $\nu$, the encoder memory spans $\nu-1$ bits, creating a code with $|\mathcal{S}| = 2^{\nu-1}$ states.  Figure~\ref{Fig::7o5::encoder} shows the encoder structure for a rate-1/2 code with $\nu=3$ and generator polynomials $g(x) = (1,\frac{1+x+x^2}{1+x^2})=(1,7/5)_{\text{oct}}$.  
The encoding process can be represented by a trellis structure (Fig.~\ref{Fig::7o5::trellis}), where each state at time $\tau$ connects to successor states at time $\tau+1$ via branches. Each branch is labeled as $u/c^{(1)}c^{(2)}$, indicating that input bit $u$ produces output bits $c^{(1)}c^{(2)}$ for that state transition. A path through the trellis uniquely determines an input sequence and its codeword.

The Viterbi algorithm performs maximum likelihood (ML) decoding of convolutional codes by finding the most probable path through the code trellis using dynamic programming \cite{viterbi}. 
\vspace{-0.5em}
\subsection{ByT5 for Language Model Priors}
\vspace{-0.3em}
To leverage linguistic context during decoding, we employ ByT5-small \cite{byt5}, a character-level language model with approximately 300M parameters. Unlike language models that use subword tokenizers, ByT5 operates directly on UTF-8 bytes, making it naturally suited for character-level processing. 

The standard ByT5 follows an \textit{encoder}-\textit{decoder} Transformer architecture. The \textit{encoder} processes context through self-attention, and the \textit{decoder} generates characters autoregressively by attending to the encoder output. In this paper, we utilize ByT5 in \textit{decoder-only} mode, where the encoder is bypassed and the decoder directly processes the character sequence. In this configuration, the decoder processes the text sequence $(t_1, \ldots, t_{j-1})$ and predicts the probability distribution over the next character $t_j$, yielding:
\begin{equation}
P(t_j | t_1, \ldots, t_{j-1};\theta),
\end{equation}
where $\theta$ denotes the model parameters. Then, the sequence probability follows from the chain rule
\begin{equation}
P(\mathbf{t}_{1:j}|\theta) = \prod_{\ell=1}^{j} P(t_\ell | t_1, \ldots, t_{\ell-1};\theta),
\end{equation}
which serves as the language model prior in our LLM-Viterbi framework. 
The ByT5-small decoder has only 4 Transformer layers, enabling efficient inference. While we employ a compact model for efficiency, our framework is model-agnostic and can generalize to any autoregressive LLMs.

\vspace{-0.5em}
\section{Proposed LLM-Viterbi Decoder}
\label{sec:proposed_decoder}
\vspace{-0.3em}


\vspace{-0.3em}
\subsection{K-best Viterbi Decoding} \label{sec:Kbest}
\vspace{-0.3em}

\begin{figure}[t]
    \centering
    \begin{subfigure}[t]{0.458\columnwidth}
        \centering
        \includegraphics[width=\textwidth]{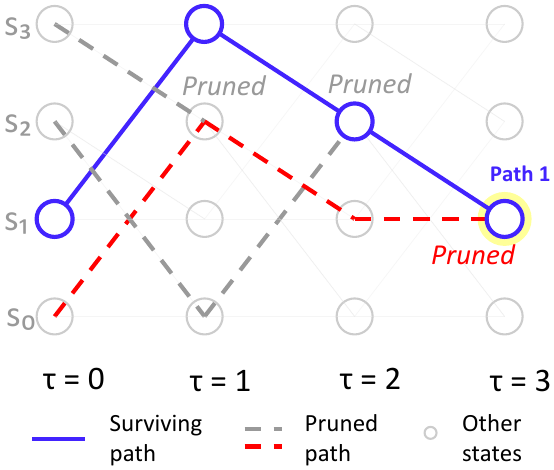}
        \caption{Standard Viterbi}
        \label{fig:standard_viterbi}
    \end{subfigure} 
    \hspace{1mm}
    \begin{subfigure}[t]{0.502\columnwidth}
        \centering
        \includegraphics[width=\textwidth]{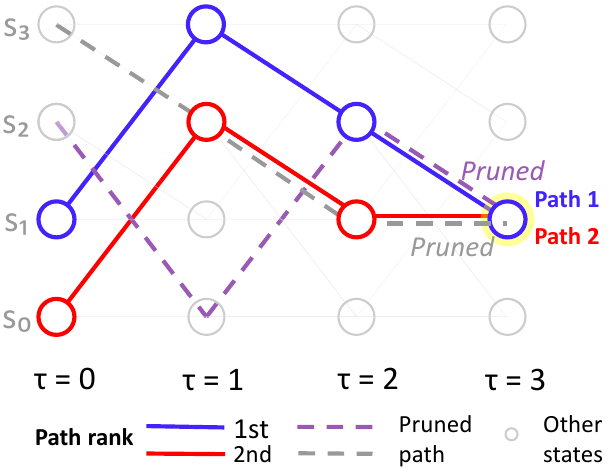}
        \caption{$K$-best Viterbi}
        \label{fig:kbest_viterbi}
    \end{subfigure}
    \vspace{-0.5em}
    \caption{Example of Viterbi and $K$-best Viterbi at a specific state $s_1$ ($\tau=3$).}
    \vspace{-1.5em}
    \label{fig:viterbi_comparison}
\end{figure}

The conventional Viterbi algorithm finds ML path on trellis. 
At each time step (i.e., stage) $\tau$ and for each state $s$, it recursively computes
\begin{equation}
M_\tau(s) = \min_{s' \in \mathcal{P}(s)} \left[ M_{\tau-1}(s') + m_{\text{branch}}(s' \to s) \right],
\end{equation}
where $\mathcal{P}(s)$ is the set of predecessor states of $s$ and $M_\tau(s)$ is the cumulative metric. The branch metric $m_{\text{branch}}(s' \to s)$ from predecessor state $s'$ to $s$ is
\begin{equation} \label{equ:branch}
m_{\text{branch}}(s' \to s) = \sum_{j=1}^{n} (y_j - x_j)^2,
\end{equation}
where $n$ is the number of coded bits per branch, $y_j$ is the received symbol, and $x_j$ is the expected transmitted symbol for  the branch $s' \to s$. 
 The algorithm retains only the single path with minimum cumulative metric at each state.


The $K$-best Viterbi algorithm extends Standard Viterbi by maintaining the $K$ best paths at each state. 
At state $s$ and time $\tau$, let $\mathcal{L}_{\tau}(s)$ denote the set of all paths arriving at $s$. Each path $p \in \mathcal{L}_{\tau}(s)$ has cumulative metric as follows. 
\begin{equation} \label{equ::extension_matrix}
M(p) = \sum_{i=1}^{\tau} m_{\text{branch}}(s_i' \to s_i),
\end{equation}
where $s_i' \to s_i$ denotes the state transition at time $i$ along the path $p$. The $K$-best selection is performed according to
\begin{equation}
\mathcal{K}_{\tau}(s) = \underset{p \in \mathcal{L}{\tau}(s)}{\text{arg\ min}_K} \ M(p),
\end{equation}
where $\text{arg\ min}_K$ returns the $K$ paths with smallest metrics. This maintains at most $K \times |\mathcal{S}|$ paths across all states at each time step. This strategy guarantees that the global top-$K$ paths are preserved within $\cup_{s \in \mathcal{S}} \mathcal{K}_{\tau}(s)$. Figure \ref{fig:viterbi_comparison} illustrate the Viterbi and K-best Viterbi decoding at a specific state.

At trellis stage $\tau$, each surviving path $p \in \mathcal{K}_{\tau}(s)$ is extended to successor states $\tilde{s} \in \mathcal{N}(s)$, where $\mathcal{N}(s)$ denotes the set of states reachable from $s$. The cumulative metric updates as
\begin{equation}
M(\mathbf{u}_{1:\tau+1}) = M(\mathbf{u}_{1:\tau}) + m_{\text{branch}}(s \to \tilde{s}),
\end{equation}
where we denote path $p$ by its decoded information bit sequence $\mathbf{u}_{1:\tau} = (u_1, u_2, \ldots, u_\tau)$, and $M(\mathbf{u}_{1:\tau}) = M(p)$. 

When $\tau = j \cdot L$ for character bit length $L=8$, the most recent $L$ bits of each path form a complete byte $\mathbf{b}_j = (u_{\tau-L+1}, \ldots, u_{\tau})$, which is mapped to character $t_j$ via ASCII encoding. Each path thus maintains a character sequence $\mathbf{t}_{1:j} = (t_1, t_2, \ldots, t_j)$.


\vspace{-0.3em}
\subsection{LLM-Assisted Path Selection}
\vspace{-0.3em}





With character sequence $\mathbf{t}_{1:j} = (t_1, t_2, \ldots, t_j)$, we can leverage a language model to evaluate the linguistic plausibility of each path. 
Let $\tau_j = j \cdot L$. For each path $p \in \mathcal{K}_{\tau}(s)$ across all states $s\in\mathcal{S}$, we maximize the posterior probability given the received signal $\mathbf{y}_{1:\tau_j}$. By Bayes' rule, we have
\begin{equation} \label{Equ::post}
\begin{split}
    P(\mathbf{u}_{1:\tau_j}|\mathbf{y}_{1:\tau_j}) &\propto P(\mathbf{y}_{1:\tau_j}|\mathbf{u}_{1:\tau_j}) \cdot P(\mathbf{u}_{1:\tau_j})\\
    & \overset{(a)}{=} P(\mathbf{y}_{1:\tau_j}|\mathbf{u}_{1:\tau_j}) \cdot P(\mathbf{t}_{1:\tau_j})
\end{split}
\end{equation}
where the evidence term $P(\mathbf{y}_{1:\tau_j})$ is omitted as it is identical across paths. Step (a) follows since  bit sequence $\mathbf{u}_{1:\tau_j}$ deterministically maps to $\mathbf{t}_{1:j}$ via ASCII.

Under AWGN channel, the channel likelihood is given by
\begin{equation}
P(\mathbf{y}_{1:\tau_j}|\mathbf{u}_{1:\tau_j}) \propto \exp\left(-\frac{M(\mathbf{u}_{1:\tau_j})}{2\sigma^2}\right),
\end{equation}
where $M(\mathbf{u}_{1:\tau_j}) = M(p)$ is the cumulative Euclidean distance. 

In conventional channel decoding, $P(\mathbf{u}_{1:\tau_j})$ is assumed uniform. However, text follows linguistic patterns, so not all character sequences are equally likely. Therefore, we proposed to estimate $P(\mathbf{t}_{1:j})$ using the fine-tuned ByT5 model (Section \ref{sec::Byt}) during the viterbi decoding. The model computes
\begin{equation}
P(\mathbf{t}_{1:j}) = \prod_{\ell=1}^{j} P(t_\ell | t_1, \ldots, t_{\ell-1}),
\end{equation}
where each conditional probability $P(t_\ell| t_1, \ldots, t_{\ell-1})$ represents the likelihood of character $t_\ell$ given all preceding characters. Particularly, $P(t_1)=P(t_1|\varnothing)$ for the first character. 

Taking logarithms of \eqref{Equ::post}, we obtain the final path score as follows.
\begin{equation} \label{equ::finalscore}
\begin{aligned}
\text{Score}(\mathbf{u}_{1:\tau_j}) \propto -\frac{M(\mathbf{u}_{1:\tau_j})}{2\sigma^2} + \log P(\mathbf{t}_{1:j}) ,
\end{aligned}
\end{equation}
where the first term reflects channel reliability and the second term reflects linguistic plausibility. Maximizing this score achieves  a maximum a posteriori (MAP) decoding of codewords by jointly optimizing over channel observations and source statistics.

Using \eqref{equ::finalscore}, a straightforward approach would evaluate all $K \times |\mathcal{S}|$ paths after each  character, select the path with the highest score, and prune all others. However, this requires  one language model inference per path per character, incurring prohibitive computational cost.

\vspace{-0.3em}
\subsection{Prefix-Based Periodic Evaluation}
\vspace{-0.3em}
To reduce computational cost while maintaining decoding performance, we introduce two strategies: (1) periodic evaluation at fixed intervals, and (2) exploiting shared ancestral paths to minimize redundant LLM calls.

\subsubsection{Periodic Evaluation} Instead of evaluating paths after every character, we perform LLM-based pruning periodically every $N$ characters. Specifically, when the decoder reaches character position $j = k \cdot N$ for $k = 1, 2, 3, \ldots$, we trigger LLM evaluation. Between evaluation points, the decoder continues $K$-best Viterbi decoding, allowing multiple paths to survive.

\subsubsection{Prefix Sharing} At evaluation position $j = k \cdot N$, multiple paths may have diverged from a common ancestral path at position $j-1 = k \cdot N - 1$. Since $K$-best Viterbi allows multiple branches from each state, a single path at position $j-1$ can spawn several successor paths that differ only in their $j$-th character. 
Specifically, paths sharing the same prefix $\mathbf{t}_{1:j-1}$ have followed an identical trajectory through the trellis up to bit position $\tau_{j-1} = (j-1) \cdot L$, and therefore share the same cumulative channel metric $M(\mathbf{u}_{1:\tau_{j-1}})$. According to \eqref{equ::finalscore}, their scores at position $j-1$ are
\begin{equation} \label{eq::prefix:score}
\text{Score}(\mathbf{u}_{1:\tau_{j-1}}) = \log P(\mathbf{t}_{1:j-1}) - \frac{M(\mathbf{u}_{1:\tau_{j-1}})}{2\sigma^2},
\end{equation}
which are identical. This enables efficient prefix-level pruning strategy. Formally, at character position $j = k \cdot N$, the pruning procedure is:
\begin{enumerate}
\item Group all $K \times |\mathcal{S}|$ surviving paths by their prefix $\mathbf{t}_{1:j-1}$. 

\item For each unique prefix $\mathbf{t}_{1:j-1}$, compute prefix score \eqref{eq::prefix:score}. 



\item Select the prefix $\mathbf{t}^*_{1:j-1}$ with the highest score.

\item Keep all paths with prefix $\mathbf{t}^*_{1:j-1}$, and prune other paths.
\end{enumerate}

Notably, all paths within the winning prefix $\mathbf{t}^*_{1:j-1}$ are retained regardless of their $j$-th character. This reduces language model calls from one per path to one per unique prefix. The $j$-th character is preserved for future evaluations.

\begin{figure}[t]
\centering
\includegraphics[width=0.8\columnwidth]{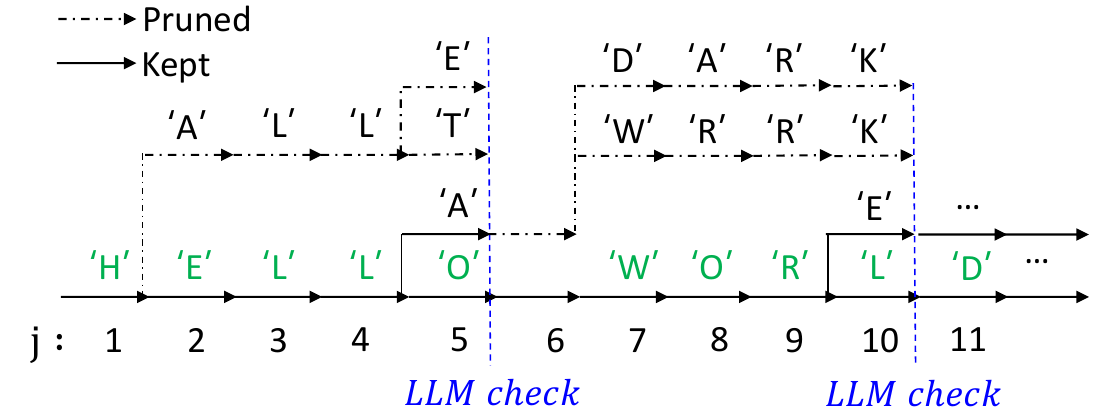}
\caption{
Prefix-based periodic pruning with $N=5$. Paths with the same prefix are grouped and evaluated together at positions $j=5, 10$.
}
\vspace{-1.5em}
\label{fig:pruning_illustration}
\end{figure}

\begin{example}

Figure~\ref{fig:pruning_illustration} illustrates the periodic pruning process with an interval of $N=5$. Between checkpoints, the $K$-best Viterbi algorithm explores multiple path extensions. Approaching $j=5$, the trellis contains several candidate sequences including the correct sequence ``HELLO'' and erroneous variants ``HELLA'', ``HALLE'', and ``HALLT''. 

At $j=5$, LLM evaluates paths by their shared prefix. The prefix ``HELL'' is identified as the most plausible context, leading to the pruning of the ``HALL'' group. At $j=10$, new paths emerge, including ``HELLO WORL'', triggering a subsequent round of LLM evaluation.
\end{example}

\subsubsection{Decoding Termination}

After LLM-based pruning at position $j = k \cdot N$, the decoder continues with $K$-best Viterbi decoding, extending the surviving paths bit-by-bit. The prefix-based LLM pruning is performed every $N$ characters until the entire message is decoded.

If the total character length is not a multiple of $N$, we perform a final evaluation. In this case, all surviving paths  are evaluated using their complete sequences according to 
\begin{equation} \label{eq::final:score}
\text{Score}(\mathbf{u}_{1:L}) = \log P(\mathbf{t}_{1:L_T}) - \frac{M(\mathbf{u}_{1:L})}{2\sigma^2}.
\end{equation}
The path with the highest score is selected as the final output.

Algorithm~\ref{alg:llm_viterbi} summarizes the complete LLM-Viterbi decoding procedure.

\begin{algorithm}[!ht]
\small
\caption{LLM-Viterbi Decoder}
\label{alg:llm_viterbi}
\begin{algorithmic}[1]
\REQUIRE Received signal $\mathbf{y}$, fine-tuned ByT5 model, $K$, $N$, $\sigma^2$
\ENSURE Decoded character sequence $\mathbf{t}$
\STATE Initialize paths at start state, $\tau \leftarrow 0$, $j \leftarrow 0$
\WHILE{not end of message}
    \STATE \textbf{// K-best Viterbi Extension}
    \FOR{each state $s \in \mathcal{S}$}
            \STATE Extend paths at $s$ to $\tilde{s} \in \mathcal{N}(s)$, update metrics via \eqref{equ::extension_matrix}
        \STATE Retain $K$ paths with smallest $M$
    \ENDFOR
    \STATE $\tau \leftarrow \tau + 1$
    \IF{$\tau \bmod 8 = 0$}
        \STATE Map latest 8 bits to character $t_j$ for each path
        \STATE $j \leftarrow j + 1$
    \ENDIF
    \STATE \textbf{// Prefix-based LLM Pruning}
    \IF{$j \bmod N = 0$ and $j > 0$}
        \STATE Group all paths by prefix $\mathbf{t}_{1:j-1}$
        \FOR{each unique $\mathbf{t}_{1:j-1}$}
             \STATE Compute score via \eqref{eq::prefix:score} using ByT5
        \ENDFOR
        \STATE Keep paths with highest-scoring prefix; prune others
    \ENDIF
\ENDWHILE
\STATE Final score evaluation of all  surviving paths according to \eqref{eq::final:score}.
\RETURN The path with highest final score.
\end{algorithmic}
\end{algorithm}

\vspace{-0.3em}
\subsection{Trade-off in Evaluation Interval}
\label{sec:trade-off}
\vspace{-0.3em}
Besides the inference overhead, evaluation interval $N$ critically affects decoding performance.

When $N$ is too small, the language model evaluates paths based on very short prefixes, leading to unreliable linguistic assessments. For example, at $N=1$ or $N=2$, comparing ``HE'' vs. ``HA'' provides insufficient context and may incorrectly favor one over the other. With larger $N$ (e.g., $N=5$), the comparison involves complete words (``HELLO'' vs. ``HALLO''), enabling more reliable linguistic evaluation and path pruning.  

When $N$ is too large, the decoder operates purely on channel metrics using $K$-best Viterbi for long periods. Correct paths may accumulate unfavorable channel metrics due to noise and be eliminated by $K$-best selection, particularly at low SNR.

The optimal $N$ balances linguistic context sufficiency against channel-based Viterbi pruning. Section~\ref{sec:results} empirically demonstrates this trade-off.

\vspace{-0.3em}
\subsection{ByT5 Model Fine-Tuning} \label{sec::Byt}
\vspace{-0.3em}

Although ByT5 is pretrained on general text, we perform full-parameter fine-tuning of the ByT5 decoder to adapt it to our LLM-Viterbi. We select 10,000 sentences with character length between 80 and 120 from the training set of the Stanford Natural Language Inference (SNLI) corpus \cite{snli}. For each sentence $\mathbf{t}_{1:L_T} = (t_1, t_2, \ldots, t_{L_T})$ of length $L_T$, we create a training sample following the standard autoregressive language modeling objective. Specifically, at each position $\ell$, the model predicts the next character $t_\ell$ given all preceding characters $t_1, \ldots, t_{\ell-1}$. In the decoder-only configuration, the ByT5 encoder receives a padding (null) token while the ByT5 decoder processes the input using causal self-attention. 


The training objective minimizes the cross-entropy loss over all character positions:
\begin{equation} 
\mathcal{L}(\theta) = - \sum_{j=1}^{L_T} \log P(t_j | t_1, \ldots, t_{j-1}; \theta),
\end{equation}
where $P(t_1|\varnothing;\theta)$ denotes the probability of the first character. 
Notably, this objective optimizes conditional probabilities for all positions $j$, making the trained model compatible with arbitrary evaluation intervals $N$ during inference.
Training uses Adam optimizer, batch size 32, learning rate $1 \times 10^{-4}$, and runs for 3 epochs. Optimization is performed by minimizing the average loss over minibatches.


\vspace{-0.5em}
\section{Experimental Results}
\label{sec:experiments}
\vspace{-0.3em}
\subsection{System Configuration}
\vspace{-0.3em}
The transmitter uses a rate-1/2 convolutional encoder with constraint length $\nu = 3$ and $5$ with generator polynomials $\mathbf{G} = (7, 5)_{\mathrm{oct}}$ and $\mathbf{G} = (35, 23)_{\mathrm{oct}}$, respectively. Source messages are drawn from the test set of SNLI corpus \cite{snli},  comprising 10,000 sentences with an approximate character length 100. Characters are encoded using 8-bit ASCII, yielding an information block length of roughly 800 bits and a coded frame length of 1600 bits. We employ ByT5-small\cite{byt5} from Hugging Face\cite{huggingface} fine-tuned as described in Section~\ref{sec::Byt}. In the proposed method, we maintain $K = 8$ paths per state, and the LLM evaluation interval is $N=5$ unless otherwise stated.

We compare the proposed method against two baselines

\subsubsection{Standard Viterbi} Conventional Viterbi decoding that selects the single path with minimum cumulative Euclidean distance.
\subsubsection{Viterbi + One-Shot LLM Correction} The received signal is first fully decoded by the Standard Viterbi algorithm. The output character sequence is then passed to a ByT5-small model for error correction.  The model is finetuned with paired data \{Viterbi output, ground truth\} at various SNRs. The  model is trained in standard encoder-decoder mode using same hyperparameters as Section \ref{sec::Byt}.

All simulations are implemented in Python using PyTorch\cite{pytorch} and the Hugging Face Transformers library\cite{huggingface}. We evaluate performance using BLER for bit-level accuracy, and SBERT (Sentence-BERT) \cite{sbert} similarity score. SBERT measures the cosine similarity between the decoded and original text, where higher scores indicate better preservation of semantic meaning. All experiments are conducted until 1000 block errors are observed for each SNR point.

\vspace{-0.3em}
\subsection{Performance Results} \label{sec:results}
\vspace{-0.3em}
\subsubsection{Block Error Rate Performance}

\begin{figure}[t]
    \centering
    \begin{subfigure}[b]{0.48\columnwidth}
        \centering
        \includegraphics[width=\linewidth]{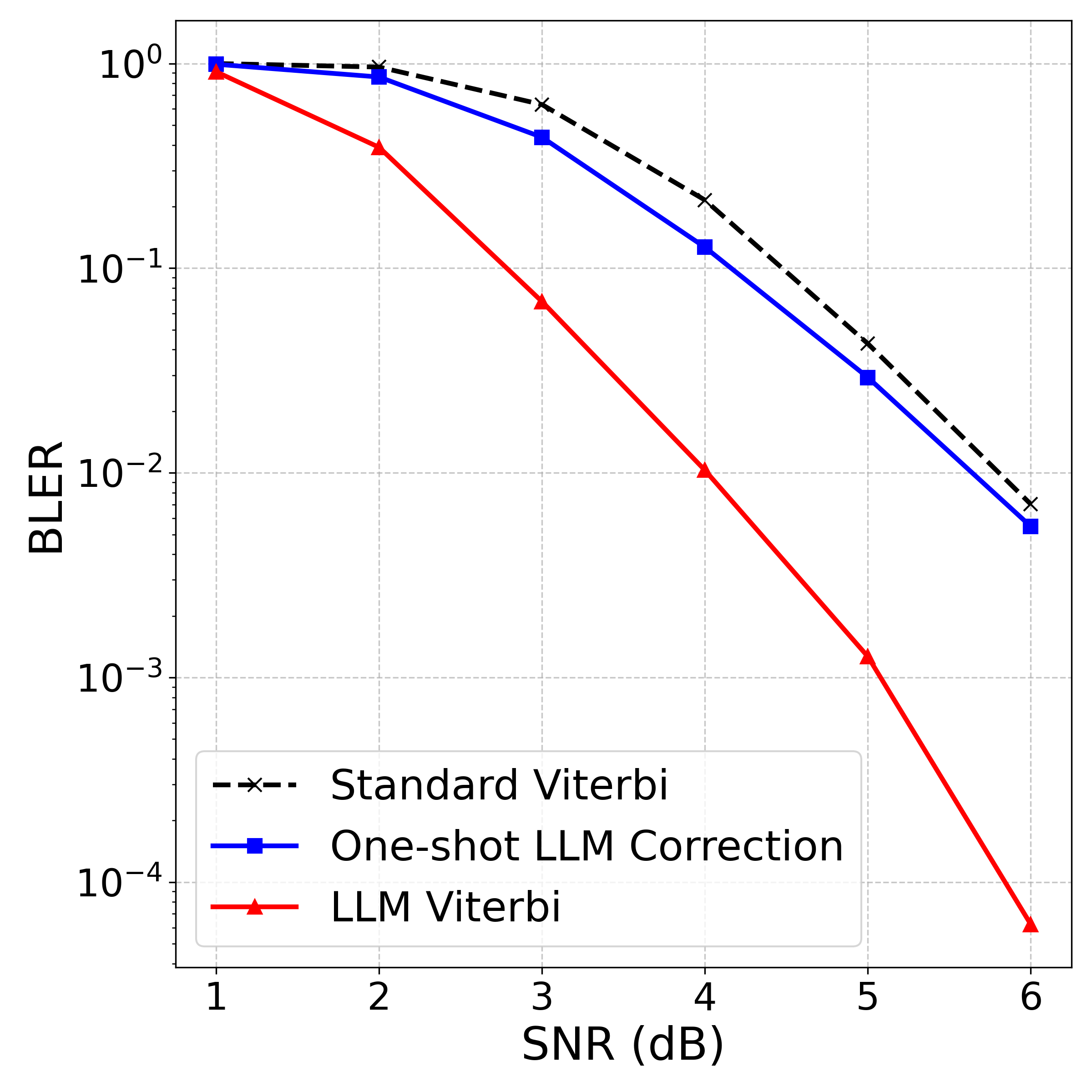}
        \vspace{-2em}
        \caption{BLER performance}
        \label{fig:bler_nu3}
    \end{subfigure}
    \hfill
    \begin{subfigure}[b]{0.48\columnwidth}
        \centering
        \includegraphics[width=\linewidth]{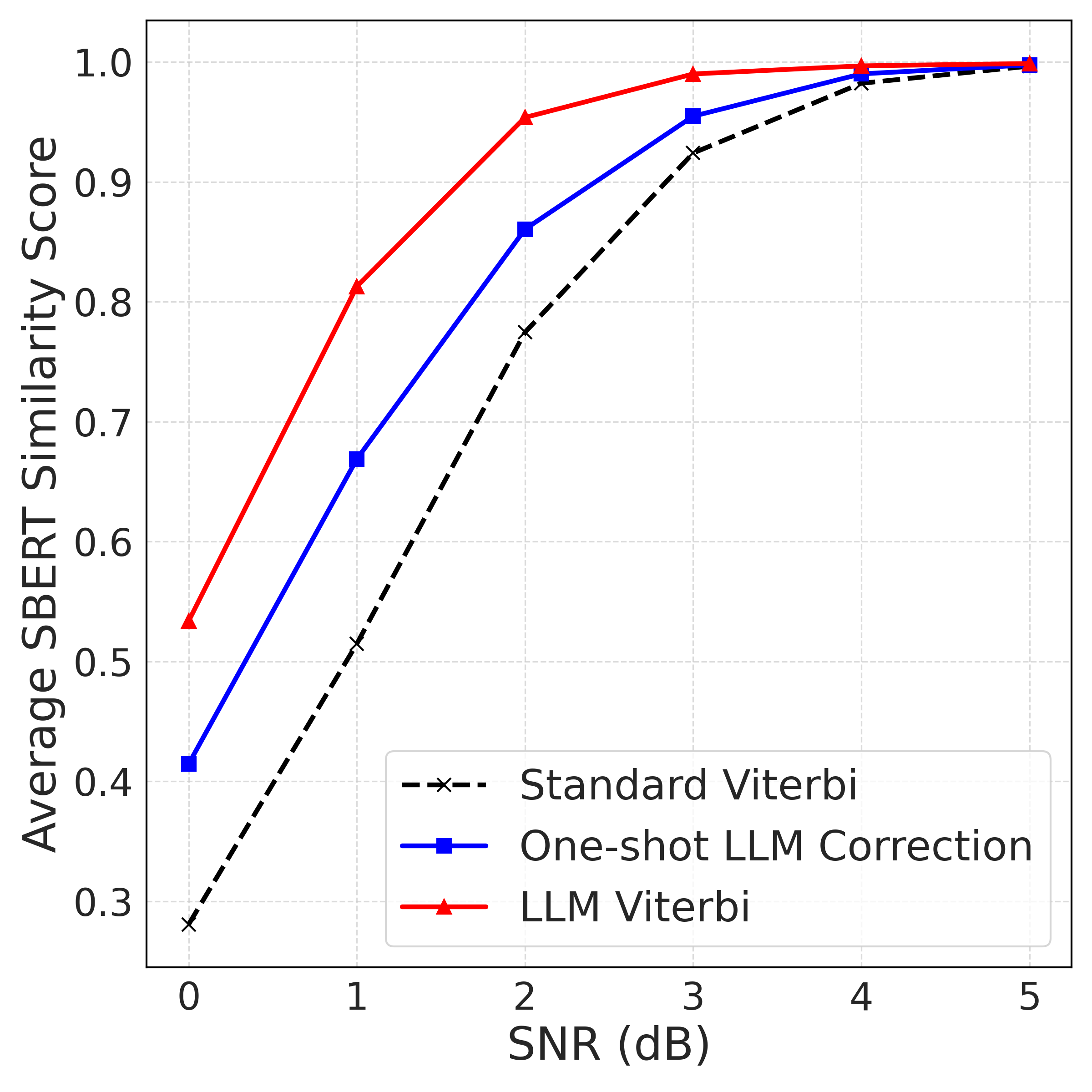}
        \vspace{-2em}
        \caption{SBERT similarity scores}
        \label{fig:sbert_nu3}
    \end{subfigure}
    \vspace{-0.5em}
    \caption{Performance comparison and semantic similarity for $\nu = 3$.}
    \label{fig:results_nu3}
    \vspace{-1em}
\end{figure}

Figure~\ref{fig:bler_nu3} compares BLER performance for $\nu = 3$. The proposed LM-Viterbi decoder achieves significant gains over both baselines across all SNR values. At SNR = 5 dB, the proposed method achieves BLER of $10^{-3}$, while Standard Viterbi and Viterbi with one-shot LLM Correction exhibit BLER of $3\times10^{-2}$ and $5\times10^{-2}$, respectively. LLM-Viterbi achieves approximately 1.5 dB over baselines. This gain stems from the joint scoring of channel reliability and linguistic context.

\begin{figure}[t]
    \centering
    \begin{subfigure}[b]{0.48\columnwidth}
        \centering
        \includegraphics[width=\linewidth]{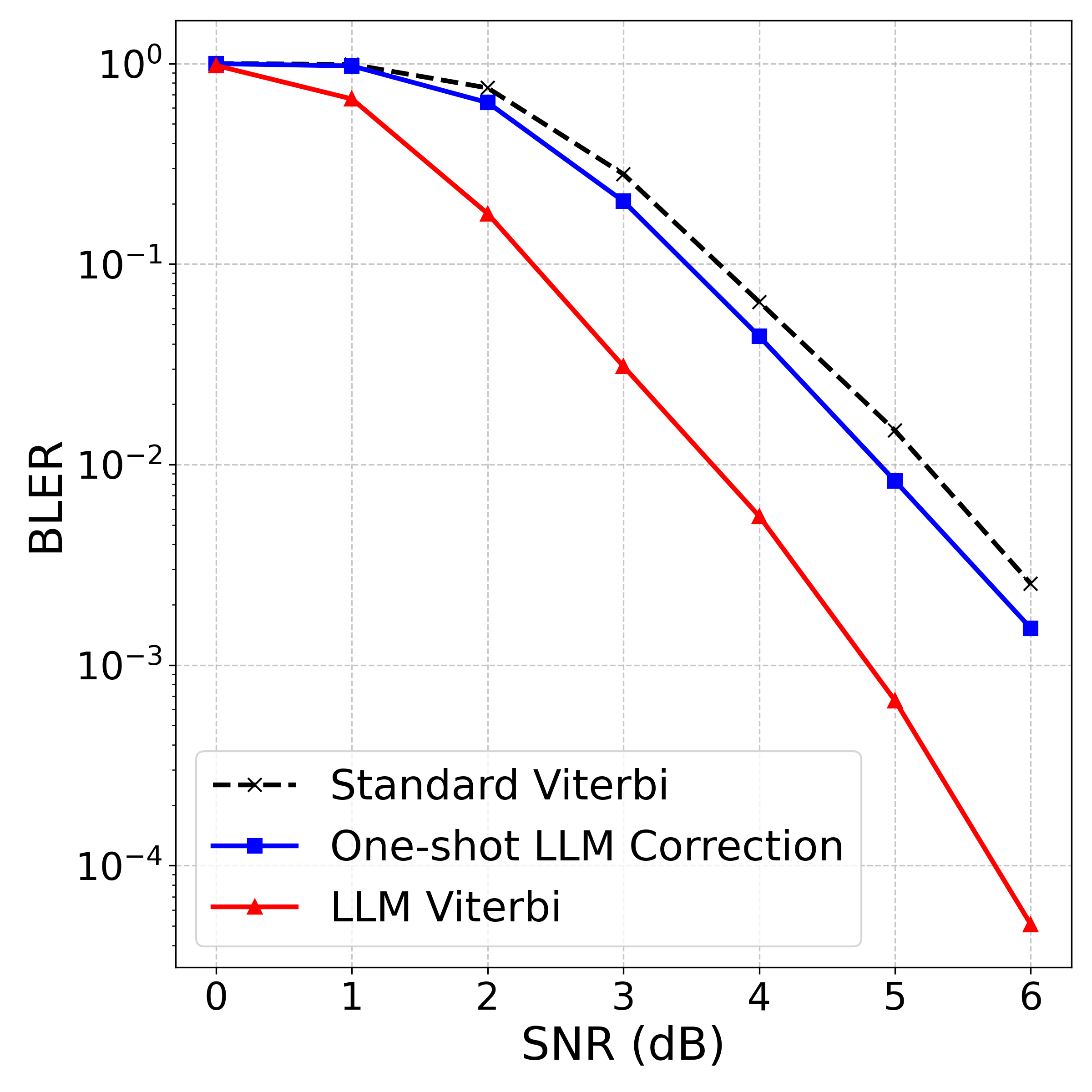}
        \vspace{-2em}
        \caption{BLER performance}
        \label{fig:bler_nu5}
    \end{subfigure}
    \hfill
    \begin{subfigure}[b]{0.48\columnwidth}
        \centering
        \includegraphics[width=\linewidth]{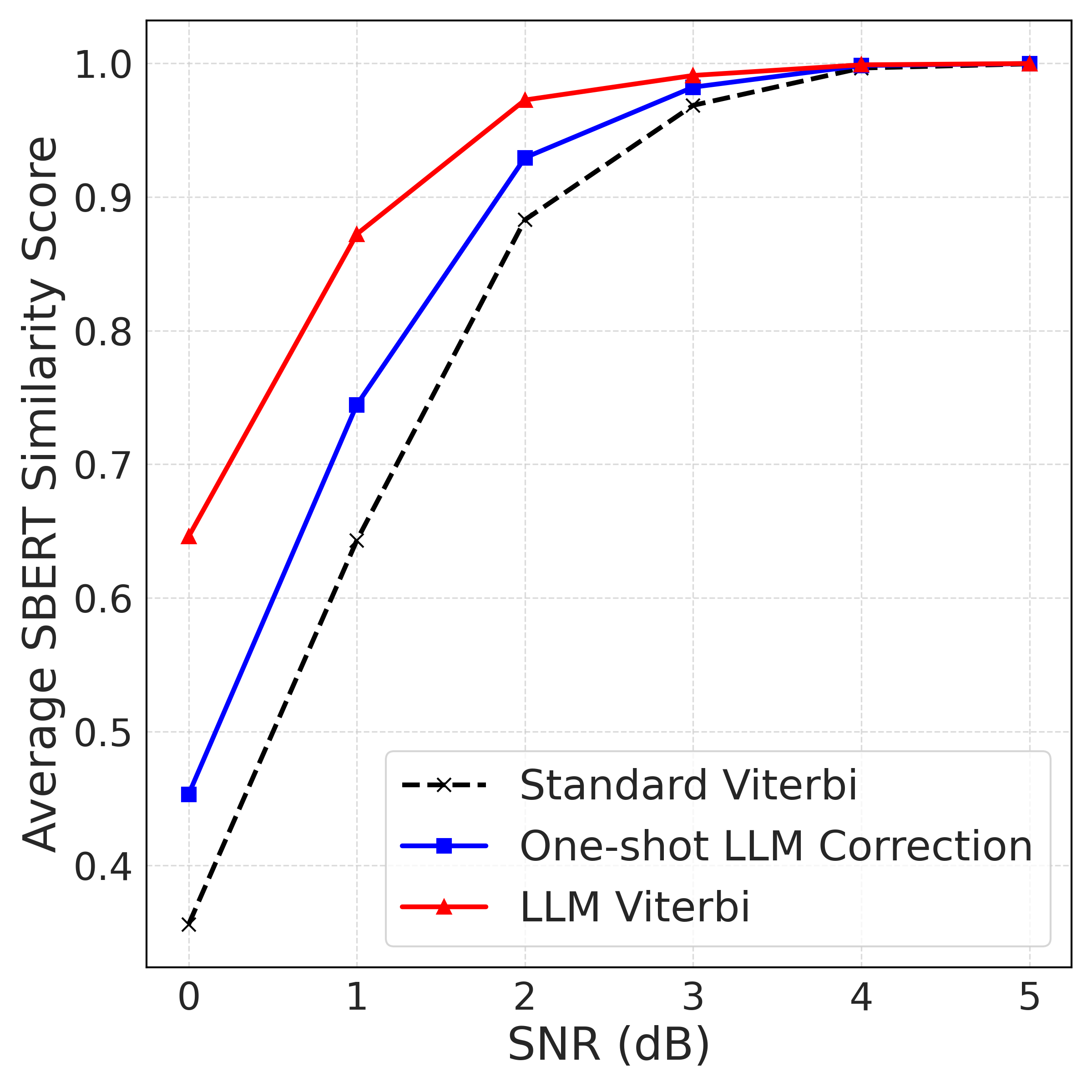}
        \vspace{-2em}
        \caption{SBERT similarity scores}
        \label{fig:sbert_nu5}
    \end{subfigure}
    \vspace{-0.5em}
    \caption{Performance comparison and semantic similarity for $\nu = 5$.}
    \vspace{-1.5em}
    \label{fig:results_nu5}
\end{figure}

For $\nu = 5$ (Fig.~\ref{fig:bler_nu5}), the longer constraint length improves error correction for all methods, while LLM-Viterbi maintains its performance advantage across the entire SNR range


\subsubsection{Semantic Similarity}

Figure~\ref{fig:sbert_nu3} shows the SBERT scores  for $\nu = 3$. The proposed LM-Viterbi decoder achieves substantially higher semantic similarity scores, especially at low SNR. At SNR = 1 dB, LM-Viterbi achieves an SBERT score of 0.82, compared to 0.66 for Viterbi with one-shot LLM Correction and 0.52 for Standard Viterbi. At SNR = 2 dB, the LM-Viterbi decoder reaches an SBERT score of 0.96, demonstrating near-perfect semantic preservation. For SNR $\geq$ 3 dB, all methods approach an SBERT of 1.0.

With $\nu = 5$ (Fig.~\ref{fig:sbert_nu5}), LLM-Viterbi continues to outperform both baselines, demonstrating that the benefits of integrating language model priors persist with more codes.

\subsubsection{Impact of LLM Evaluation Interval $N$}

\begin{figure}[t]
\centering
\includegraphics[width=0.85\columnwidth]{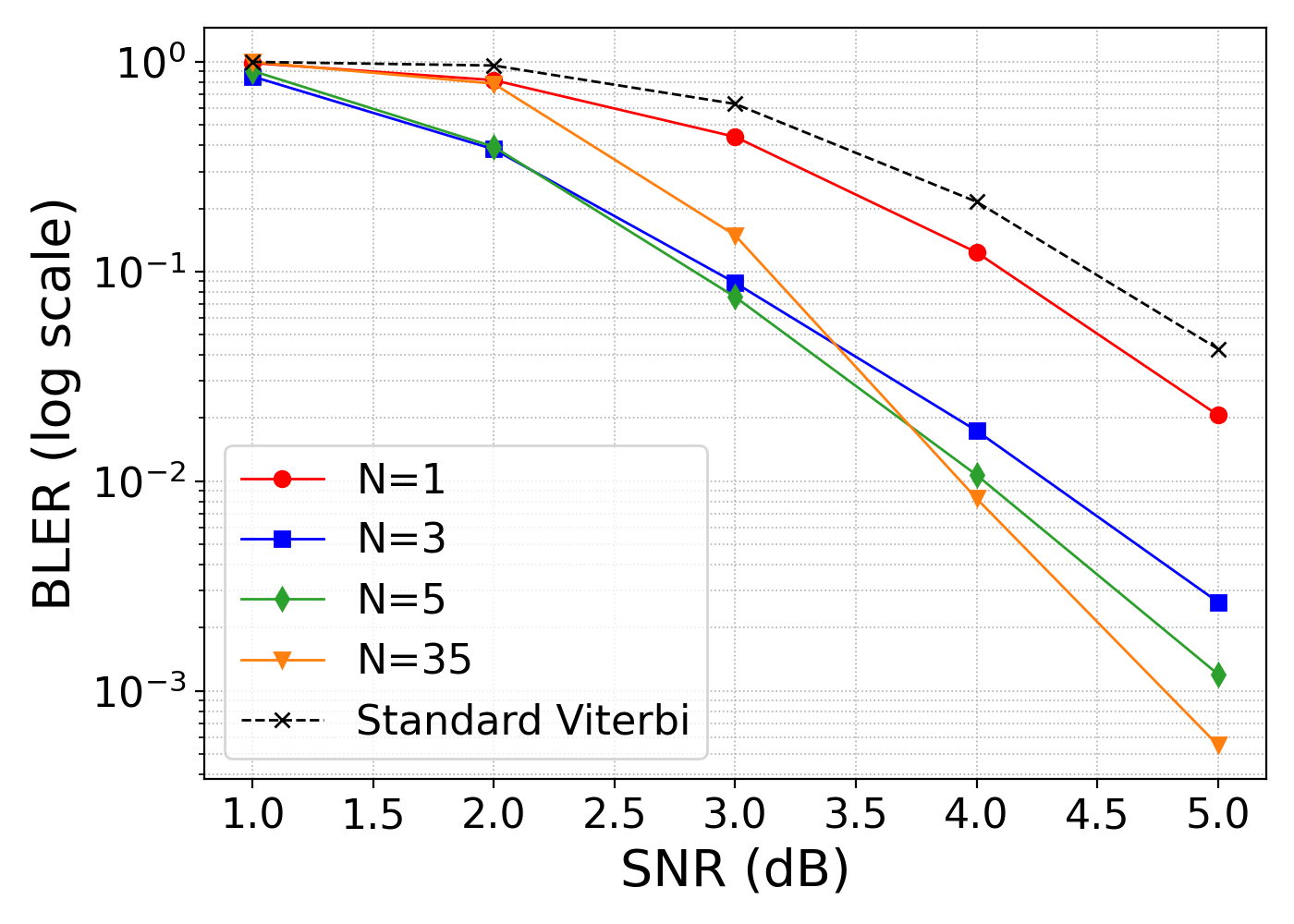}
\vspace{-1em}
\caption{BLER performance with different LLM evaluation intervals $N=1,3,5$ and $35$ for constraint length $\nu = 3$.}
\label{fig:different_N}
\vspace{-0.5em}
\end{figure}

We evaluate the proposed method with different evaluation intervals $N \in \{1, 3, 5, 35\}$ for constraint length $\nu = 3$. Fig.~\ref{fig:different_N} shows the BLER performance across SNR values.
At $N=1$, BLER performance is poor across all SNRs due to premature pruning based on insufficient context, as discussed in Section~\ref{sec:trade-off}. The decoder makes greedy decisions at each character, eliminating paths that may become plausible with additional observations.


Compared to $N=1$, $N=3$ shows modest improvement across all SNRs. $N=5$ achieves strong performance consistently across the entire SNR range, demonstrating robust decoding. However, $N=35$ exhibits notable performance degradation at low SNRs (1-3 dB). With infrequent LLM evaluation, the decoder relies predominantly on channel metrics through extended periods of $K$-best Viterbi selection, causing incorrect paths to accumulate favorable channel metrics while correct paths may be eliminated. At high SNRs, $N=35$ performs well as reliable channel observations allow language model evaluation based on longer context. This trade-off aligns with the discussion in Section~\ref{sec:trade-off}.



\vspace{-0.3em}
\subsection{Decoding Latency}
\vspace{-0.3em}
Table~\ref{tab:decoding_time} compares the average decoding time per message block. We also include code $(371,247)_{\mathrm{oct}}$ with constraint length $\nu = 8$ for a complete comparison. For constraint length $\nu = 3$, the proposed LLM-Viterbi decoder (315 ms) is approximately 10$\times$ slower than standard Viterbi (30 ms) but achieves comparable latency to the one-shot LLM approach (354 ms). The increased latency stems from (1) periodic LLM evaluation every $N = 5$ characters, and (2) maintaining $K$ paths at each trellis state instead of a single path.

For $\nu = 5$ and $\nu = 8$, the decoding time of LLM-Viterbi is substantially higher than both baselines. This is attributed to the exponential growth in trellis states. the number of candidate paths becomes $K \times |\mathcal{S}| = 8 \times 16 = 128$ for $\nu = 5$ and $K \times |\mathcal{S}| = 8 \times 128 = 1024
$ for $\nu = 8$, compared to only 32 paths for $\nu = 3$. Evaluating these paths every 5 characters introduces significant computational overhead.

We acknowledge that integrating language model priors substantially increases decoding latency compared to standard Viterbi. However, this computational cost yields significant performance gains in both BLER and semantic quality, as demonstrated in Section~\ref{sec:results}. Future work may explore optimizations for reduced inference overhead.

\begin{table}[t]
\centering
\caption{Average decoding time comparison (ms per block)}
\label{tab:decoding_time}
\vspace{-0.3em}
\begin{tabular}{|l|c|c|c|}
\hline
\multirow{2}{*}{Method} & \multicolumn{3}{c|}{Constraint Length} \\
\cline{2-4}
 & $\nu = 3$ & $\nu = 5$ & $\nu = 8$  \\
\hline
Standard Viterbi         & 30  & 99  & 721  \\
Viterbi + one-shot LLM Correction & 354 & 430  & 1050\\
Proposed LLM-Viterbi     & 315 & 1021 & 7888\\
\hline
\end{tabular}
\vspace{-1.5em}
\end{table}

\vspace{-0.5em}
\section{Conclusion}
\vspace{-0.5em}
This paper proposed an LLM-Viterbi decoder that integrates language model priors into Viterbi decoding with $K$-best paths for text transmission over noisy channels.  By combining channel reliability metrics with linguistic scores from a fine-tuned ByT5 model, the decoder selects paths jointly based on channel observations and semantic coherence. Experimental results demonstrate significant improvements in both BLER and Semantic similarity. 


\bibliographystyle{IEEEtran}
\bibliography{references}

\end{document}

%% file: figures/7o5_encoder.tex
       \begin{tikzpicture}[auto, node distance=2cm, >=latex']
            \node [inner sep=-1pt] (oplus) at (0, 0) {$\oplus$};
            \node [inner sep=-1pt, above right=0.6cm and 1.45cm of oplus] (oplus1) {$\oplus$};
            \node [inner sep=-1pt, above right=0.6cm and 2.81 cm of oplus] (oplus2) {$\oplus$};
            \node [draw, rectangle, minimum height=0.74cm, minimum width=0.9cm, right of=oplus, node distance=1cm] (S1) {$1$};
            \node [draw, rectangle, minimum height=0.74cm, minimum width=0.9cm, right of=S1, node distance=1.4cm] (S2) {$2$};
            
            \node [left of=oplus, node distance=0.7cm, inner sep=2pt] (input) {$u$};
            \node [right of=oplus2, node distance=0.5cm, inner sep=0pt] (output2) {$c^{(2)}$};
            \node [above of=output2, node distance=0.7cm,inner sep=0pt] (output1) {$c^{(1)}$};
            
            \draw [->] (input) -- (oplus);
            \draw [->] (oplus) -- (S1) ;
            \draw [->] (S1) -- (S2);
            \draw [->] (S2) -- ++(0.63,0)  |- ($(oplus.west) + (+0.5,-0.75)$) -| (oplus);
            \draw [->] (oplus2) -- (output2);
            \draw [->] (S2.east) -|  (oplus2.south);
            \draw [->] (S1)   -|  (oplus1.south);
            \draw [->] (oplus)-- ++(0.25,0) |-   (oplus1.west);
            \draw [->] (oplus1)-- (oplus2);
            \draw [->] (input) -- ++(0.4,0)  |- (output1);
        \end{tikzpicture}    

%% file: figures/7o5_trellis.tex
        \begin{tikzpicture}[
                node distance=0.6cm,
                state/.style={circle, draw, minimum size=0.5cm,inner sep=0pt},
                edge/.style={->, >=stealth', shorten >=1pt, auto, node distance=2cm, semithick}
            ]
            
            \node[state] (S0) {$00$};
            \node[state, below of=S0] (S1) {$01$};
            \node[state, below of=S1] (S2) {$10$};
            \node[state, below of=S2] (S3) {$11$};
            
            \node[state, right of=S0,node distance=3cm] (S0next) {$00$};
            \node[state, below of=S0next] (S1next) {$01$};
            \node[state, below of=S1next] (S2next) {$10$};
            \node[state, below of=S2next] (S3next) {$11$};
            
            \draw[edge] (S0) -- node[above, pos=0.15,sloped, yshift=-1mm, font=\tiny] {$0/00$} (S0next);
            \draw[edge] (S0) -- node[above, pos=0.15,sloped, yshift=-1mm, font=\tiny] {$1/11$} (S2next);
            \draw[edge] (S1) -- node[above, pos=0.15,sloped, yshift=-1mm, font=\tiny] {$0/00$} (S2next);
            \draw[edge] (S1) -- node[above, pos=0.15,sloped, yshift=-1mm, font=\tiny] {$1/11$} (S0next);
            \draw[edge] (S2) -- node[above, pos=0.15,sloped, yshift=-1mm, font=\tiny] {$0/01$} (S1next);
            \draw[edge] (S2) -- node[above, pos=0.15,sloped, yshift=-1mm, font=\tiny] {$1/10$} (S3next);
            \draw[edge] (S3) -- node[above, pos=0.15,sloped, yshift=-1mm, font=\tiny] {$0/01$} (S3next);
            \draw[edge] (S3) -- node[above, pos=0.15,sloped, yshift=-1mm, font=\tiny] {$1/10$} (S1next);
            
            \node[above of=S0, node distance=0.4cm, font=\footnotesize] {Time $\tau$};
            \node[above of=S0next, node distance=0.4cm, font=\footnotesize] {Time $\tau+1$};
            
        \end{tikzpicture}